\documentclass[epj]{svjour}
\usepackage{graphicx}
\usepackage{amsmath}
\usepackage{bm}
\usepackage{amssymb}
\usepackage{latexsym}
\usepackage{dcolumn}
\usepackage{captcont}
\usepackage{url}

\newcolumntype{d}{D{.}{.}{10}}
\begin{document}
\title{Hyperfine Quenching of the $4s4p\ ^{3}P_{0}$  Level in Zn-like Ions}
\author{J.\ P.\ Marques\inst{1} \and
        F.\ Parente\inst{2} 
        \and 
        P.\ Indelicato\inst{3}
%
}                     
\offprints{J.\ P.\ Marques}          
\institute{Centro de F{\'\i}sica At{\'o}mica e Departamento F{\'\i}sica, 
Faculdade de Ci{\^e}ncias, Universidade de Lisboa, \\
Campo Grande, Ed. C8, 1749-016 Lisboa, Portugal, 
\email{jmmarques@fc.ul.pt}
\and
Centro de F{\'\i}sica At{\'o}mica da Universidade de Lisboa e Departamento F{\'\i}sica da 
Faculdade de Ci{\^e}ncias e Tecnologia da Universidade Nova de Lisboa, Monte da Caparica, 2825-114 Caparica,  Portugal,
\email{facp@fct.unl.pt}
\and
Laboratoire Kastler Brossel,
\'Ecole Normale Sup\' erieure; CNRS; Universit\' e P. et M. Curie - Paris 6\\
Case 74; 4, place Jussieu, 75252 Paris CEDEX 05, France,
\email{paul.indelicato@spectro.jussieu.fr}
}
\date{Received: \today / Revised version: date}
%

\abstract{
In this paper, we used the multiconfiguration Dirac-Fock method to compute with high precision the influence of the hyperfine interaction on the $[Ar]3d^{10} 4s4p\ ^3P_0$ level lifetime in Zn-like ions for stable and some quasi-stable isotopes of nonzero nuclear spin between $Z=30$ and $Z=92$. The influence of this interaction on the $[Ar]3d^{10} 4s4p\ ^3P_1 - [Ar]3d^{10} 4s4p\ ^3P_0$ separation energy is also calculated for the same ions.
} 
\maketitle
%

\section{Introduction}
\label{intro}
It has been found before that the hyperfine interaction plays a fundamental
role in the lifetimes and energy separations of the $^3P_0$ and $^3P_1$
levels of the configurations $1s2p$ in He-like~\cite{mohr1,ind1,rm1,dun1,simi,abouss1,volot,toleikis04,johnson}, [He]$2s2p$ in Be-like~\cite{jm1,brage98,brage02}, and [Ne]$3s3p$ in Mg-like~\cite{brage98,jm2} ions, and also in $3d^4 J=4$ level in the Ti-like ions~\cite{parente94}.

In the He-like ions, in the region $Z\approx 45$, these two levels undergo a level 
crossing~\cite{joh1} and are nearly degenerate due to the electron-electron magnetic interaction, which
leads to a strong influence of the hyperfine interaction on the energy
splitting and on the $^3P_0$ lifetime for isotopes with nonzero nuclear
spin.  In Be-like, Mg-like and Zn-like ions a level crossing of the $^{3}P_{0}$ 
and $^{3}P_{1}$ levels has not been found~\cite{jm4}, but hyperfine interaction still has
strong influence on the lifetime of the $^3P_0$ metastable level and on the energy splitting, for isotopes with nonzero nuclear spin.

Until recently, laboratory measurements of  atomic mestastable states hyperfine quenching have been performed only for He-like systems, for $Z=28$ to $Z=79$~\cite{rm1,dun1,simi,toleikis04,indelicato92,dunford93,birkett93,simi94}. Hyperfine-induced transition lines in Be-like systems have been found in the planetary nebula NGC3918~\cite{brage02}. Measured values of transition probabilities were found to agree with computed values~\cite{brage98,jm3}.

Divalent atoms are being investigated, both theoretically and experimentally, in order to investigate the possible use of the  hyperfine quenched $^3P_0$ metastable state for ultraprecise optical clocks and trapping experiments~\cite{porsev04}.

Recently,  dielectronic recombination rate coefficients were measured for three isotopes of Zn-like Pt$^{48+}$ in the Heidelberg heavy-ion storage ring TSR~\cite{schippers05}. It was suggested that hyperfine quenching 
of the $4s4p\ ^{3}P_{0}$ in isotopes with non-zero nuclear spin could explain the differences  detected in the observed spectra.

In this paper we extend our previous calculations~\cite{ind1,jm1,jm2} to the
influence of the hyperfine interaction on the $1s^{2} 2s^{2} 2p^{6} 3s^{2} 3p^{6} 3d^{10} 4s4p$
levels in Zn-like ions.

\begin{figure*}
	\centering
		\includegraphics[width=6.5cm]{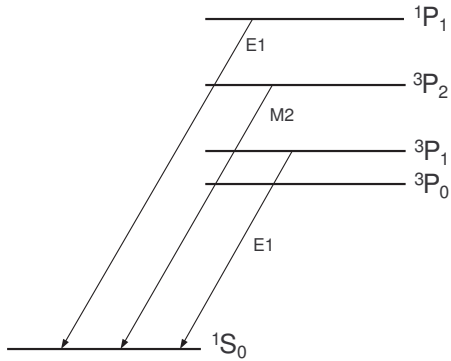}
	\caption{Energy level and transition scheme for Zn-like ions (not to scale).}
	\label{fig1}
\end{figure*}

In Fig.~\ref{fig1} we show the  energy level 
scheme, not to scale, of these ions. The $^3P_0$ level is a metastable level;
one-photon transitions from this level to the ground state are forbidden,
and multiphoton transitions have been found to be negligible in similar 
systems, so the same behavior can be expected for Zn-like ions.
Therefore, in first approximation, we will consider the
lifetime of this level as infinite.  The energy separation between the $4\ ^3P_0$ 
and $4 \ ^3P_1$ levels is small for $Z$ values around the neutral Zn atom and
increases very rapidly with $Z$. Hyperfine interaction is expected to
have a strong influence on the energy splitting and on the
$4 \ ^3P_0 $ lifetime for isotopes with nonzero nuclear spin.  

	The different steps of this calculation are described in
Refs.\ ~\cite{ind1,cheng85}. Here we will emphasize only the fundamental
topics of the theory and the characteristic features of
Zn-like systems.

In this work we used the multi-configuration Dirac-Fock code of 
Desclaux and Indelicato~\cite{descl01,indel01,mcdf} to  evaluate, completely \textit{ab initio} the $4s4p$ fine structure energies and transition probabilities.

Several terms, such as the nonrelativistic ($J$ independent) contribution to the
correlation, are the same for all levels. In this calculation Breit interaction
is included, as well as radiative corrections, using the method described, for instance, in Ref.~\cite{sant1}~
and references therein. The MCDF method, in principle, allows for
 precise calculations because it can include most of the
 correlation relatively easily, i.\ e., with a small number of
 configurations.  Here, correlation is important in the
 determination of transition energies to the ground state, which
 are used in the calculation of transition probabilities.
We found that the largest effect is obtained by using [Ar]$3d^{10} 4s^2$ and $4p^2$ 
as the configuration set for the ground state.  For the excited states we included
all configuration state functions (CSF) originated  from [Ar]$3d^{10} 4s4p$, $4p4d$,
and $4d4f$ (which is usually defined as intrashell correlation), because in second-order perturbation
theory the dominant energy difference denominators correspond to these configurations.
Correlation originating from interaction with the Ar-like core has been neglected. In particular we included, in a test calculation, spin-polarization from $s$ subshells and found a negligible influence in both energies and transition probabilities. 
We note that the fine structure energy separation, $\Delta E_{0;fs}=E_{^3P_1}-E_{^3P_0}$, which is
the important parameter in the calculation of the hyperfine quenching, is not very
sensitive to correlation, similarly to what we found for systems with smaller
number of electrons~\cite{jm1,jm2}. The same set of CSF have been used for energy,
transition probabilities, and for the calculation of the hyperfine matrix elements.

All energy calculations
are done in the Coulomb gauge for the retarded part of the
electron-electron interaction, to avoid spurious contributions
(see for example Refs.\ ~\cite{gorceix88,lin90}).  The lifetime calculations
are all done using exact relativistic formulas.  The length gauge
has been used for all transition probabilities.

\section{Relativistic calculation of hyperfine contribution to
fine structure splitting and to transition probabilities}
\label{sec:Relcalc}
In the case of a nucleus with nonzero spin, the hyperfine
interaction between the nucleus and the electrons must be taken
into account.  The correspondent Hamiltonian can be written as
\begin{equation}
H_{\textrm{hfs}} = \sum_{k} {\bf M}^{(k)} {\bf \cdot T}^{(k)}, 
\label{eq:Hhfs}
\end{equation}
where $ {\bf M}^{(k)}$and $ {\bf T}^{(k)}$ are spherical tensors of rank
$k$, representing, respectively, the nuclear and the atomic parts of the
interaction.  As in the case of He-like, Be-like and Mg-like ions, the only sizable
contribution from Eq.~(\ref{eq:Hhfs}) is the magnetic dipole term
($k=1$).  The hyperfine interaction mixes states with the same $F=J+I$ values. 
In our case, we are interested in the $^3P_0$ level, so $J=0$ and $F=I$. 
The contribution of this interaction for the total energy has
been evaluated through the diagonalization of the matrix:

\begin{align}
&H_{\textrm{tot}} =  \nonumber \\
&\left[ \begin{array}{ccc}
E_0 + \frac{1}{2}i \Gamma_0  + W_{0,0} & W_{0,1} & W_{0,2}\\ 
W_{1,0} & E_1+ \frac{1}{2} i \Gamma_1 + W_{1,1} & W_{1,2} \\ 
W_{2,0} & W_{2,1} & E_2 + \frac{1}{2}i \Gamma_2  + W_{2,2} \\
\end{array} \right]  \
\label{eq:H}
\end{align}

Here, $E_f$ is the unperturbed level energy and $\Gamma_f$ is the
radiative width of the unperturbed level ($f=0, 1, 2$ stands,
respectively, for $^3P_0, ^3P_1, ^1P_1$).  In reality there is a 
fourth level, $^3P_2$, which we included in all calculations but was found unnecessary, 
because the influence of this level in the  $^3P_0$ lifetime is negligible. 
This can be explained by the large energy separation between the $^3P_2$ and $^3P_0$
levels and also because the probability of the allowed M2 transition
$4s4p\ ^3P_2 \rightarrow 4s^2\ ^1S_0$ is many orders of magnitude smaller than those of
the E1 transitions $4s4p\ ^3P_1 \rightarrow 4s^2\ ^1S_0$ and $4s4p\ ^1P_1 \rightarrow 4s^2\ ^1S_0$.
Also, the magnetic dipole hyperfine matrix element between the $^3P_2$ and $^3P_0$ levels is very small.
This also leads to a negligible  influence of the $^3P_2$ level on the $^3P_1 - ^3P_0$ separation
energy. This is consistent with the results of Plante e Johnson~\cite{Johnson97}, who found that the magnetic
quadrupole term of the hyperfine interaction
affects the $^3P_2$ level only at  high $Z$.
 The influence of the $^1P_1$ level, however,  must be taken into account, specially for light nuclei, 
because the large energy separation between $^1P_1$ and $^3P_0$ levels is compensated by the much shorter lifetime of the $^1P_1$ level. The hyperfine matrix element
\[
W_{f,f'}=W_{f',f} =\langle  \textrm{[Ar]}\ 3d^{10} 4s4p\ f|H_{hfs}|\textrm{[Ar]}\ 3d^{10} 4s4p\ f'\rangle 
\]
may be written as
\begin{equation}
W_{J_1,J_2}=\langle I,J_1,F,M_F|{\bf M}^{(1)}{\bf \cdot T}^{(1)}|I,J_2,F,M_F \rangle,
\end{equation}
where $I$ is the nuclear spin and $F$ the total angular
momentum of the atom, and may be put in the form:
\begin{align}
	W_{J_1,J_2} &=
	(-1)^{I + J_1 + F} \left\{ 
	\begin{array}{ccc}
I & J_1 & F \nonumber \\
J_2 & I & 1
\end{array}
\right\}
\\
&\times \langle I||{\bf M}^{(1)}||I \rangle \langle J_1||{\bf
T}^{(1)}||J_2 \rangle.
\end{align}
The $6j$ symbol leads to $W_{0,0}=0$.  Also the nuclear magnetic moment
$\mu_I$ in units of the nuclear magneton $\mu_N$ may be defined by
\begin{equation}
 \mu_I \mu_N = \langle I||{\bf M}^{(1)}||I \rangle
\left( \begin{array}{ccc} I & 1 & I \\ -I & 0 & I \end{array} \right),
\end{equation}
with $\mu_N = e h/2 \pi m_p c$.

	The electronic matrix elements were evaluated on the basis set 
$|^3P_0 \rangle$, $|^3P_1 \rangle$, $|^1P_1 \rangle$
with all intrashell correlation included.

The final result is then obtained by a diagonalization of the $3 \times 3$
matrix in Eq.  (\ref{eq:H}), the real part of each eigenvalue being the
energy of the correspondent level and the imaginary part its lifetime. 

\section{Results and discussion}
\label{sec:Res}

The MCDF method  has been
used to evaluate the influence of the hyperfine interaction on
the [Ar]$3d^{10} 4s4p\ ^3P_0$, $^3P_1$ and $^1P_1$ levels for all $Z$
values between 30 and 92 and for all stable and some quasi-stable
isotopes of nonzero nuclear spin.  Contributions from other levels
have been found to be negligible. A detailed list of the contributions
to the the theoretical $^3P_0$, $^3P_1$ and $^1P_1$ level energies is
presented in Table~\ref{energies}, for $Z=36, 54$, and $82$.

\begin{table*}[h]
\begin{center}
\caption{Contribution to the energy of the $4s4p ^3P_0, ^3P_1$ and $^1P_1$ levels (in eV). }
\label{energies}

\begin{tabular}{lddd}\hline \hline
  	&	\multicolumn{1}{c}{$^3P_0$}	&	\multicolumn{1}{c}{$^3P_1$}	&	\multicolumn{1}{c}{$^1P_1$}	\\
\hline
       & \multicolumn{3}{c}{$Z=36$} \\
Coulomb$^{\dagger}$	&	-75616.613	&	-75616.265	&	-75609.264	\\
Magnetic$^{\dagger}$	&	42.785	&	42.779	&	42.765	\\
Retardation (order $\omega^2$)$^{\dagger}$	&	-4.095	&	-4.095	&	-4.094	\\
Retardation ($ > \omega^2$)	&	-0.197	&	-0.197	&	-0.197	\\
Self-energy (SE)	&	31.358	&	31.363	&	31.366	\\
Self-energy screening	&	-2.735	&	-2.740	&	-2.743	\\
VP [ $\alpha (Z\alpha)$]correction to e-e interaction	&	0.033	&	0.033	&	0.033	\\
Vacuum Polarization $\alpha (Z\alpha)^3$ +  $\alpha^2 (Z\alpha)$	&	0.010	&	0.010	&	0.010	\\
2nd order (SE-SE + SE-VP + S-VP-E)$^{\ddagger}$	&	-0.031	&	-0.031	&	-0.031	\\
Recoil	&	-0.003	&	-0.003	&	-0.003	\\
Relativistic Recoil$^{\sharp}$	&	0.009	&	0.009	&	0.009	\\
Total energy	&	-75549.478	&	-75549.137	&	-75542.148	\\
      & \multicolumn{3}{c}{$Z=54$} \\
Coulomb$^{\dagger}$	&	-195348.063	&	-195344.974	&	-195317.667	\\
Magnetic$^{\dagger}$	&	171.169	&	171.142	&	170.943	\\
Retardation (order $\omega^2$)$^{\dagger}$	&	-17.418	&	-17.418	&	-17.417	\\
Retardation ($ > \omega^2$)	&	-1.711	&	-1.713	&	-1.731	\\
Self-energy (SE)	&	127.610	&	127.615	&	127.669	\\
Self-energy screening	&	-9.166	&	-9.170	&	-9.205	\\
VP [ $\alpha (Z\alpha)$]correction to e-el interaction	&	0.124	&	0.124	&	0.124	\\
Vacuum Polarization $\alpha (Z\alpha)^3$ +  $\alpha^2 (Z\alpha)$	&	0.242	&	0.242	&	0.242	\\
2nd order (SE-SE + SE-VP + S-VP-E)$^{\ddagger}$	&	-0.253	&	-0.253	&	-0.253	\\
Recoil	&	-0.012	&	-0.012	&	-0.012	\\
Relativistic Recoil$^{\sharp}$	&	0.042	&	0.042	&	0.042	\\
Total energy	&	-195077.437	&	-195074.376	&	-195047.265	\\
      & \multicolumn{3}{c}{$Z=82$} \\
Coulomb$^{\dagger}$	&	-513070.812	&	-513061.827	&	-512876.005	\\
Magnetic$^{\dagger}$	&	710.586	&	710.511	&	708.897	\\
Retardation (order $\omega^2$)$^{\dagger}$	&	-72.655	&	-72.654	&	-72.655	\\
Retardation ($ > \omega^2$)	&	-14.666	&	-14.668	&	-14.965	\\
Self-energy (SE)	&	600.177	&	600.173	&	600.249	\\
Self-energy screening	&	-38.198	&	-38.203	&	-38.200	\\
VP [ $\alpha (Z\alpha)$]correction to e-e interaction	&	0.572	&	0.572	&	0.570	\\
Vacuum Polarization $\alpha (Z\alpha)^3$ +  $\alpha^2 (Z\alpha)$	&	4.192	&	4.192	&	4.189	\\
2nd order (SE-SE + SE-VP + S-VP-E)$^{\ddagger}$	&	-2.975	&	-2.975	&	-2.974	\\
Recoil	&	-0.051	&	-0.051	&	-0.051	\\
Relativistic Recoil$^{\sharp}$	&	0.191	&	0.191	&	0.191	\\
Total energy	&	-511883.638	&	-511874.738	&	-511690.755	\\
\hline
\end{tabular}
\end{center}
$^{\dagger}$ Contains the Uheling potential contribution to all order and all order Breit interaction.\newline
$^{\ddagger}$ Calculated using the results of Ref.~\cite{Yerokhin03,Yerokhin05,Yerokhin05a}.\newline
$^{\sharp}$ The formulas and definitions used to evaluate this term are Appendix A of Ref.~\cite{Mohr05}
\end{table*}

In Table~\ref{matr_el} we present, for all possible
values of the nuclear spin $I$, $Z$, and the mass number $A$, the diagonal and
off-diagonal hyperfine matrix elements $W_{i,j}$, and the nuclear magnetic moment, $\mu_I$~\cite{raghavan89},  in nuclear magneton units. The indexes 0, 1, and 2 in the hyperfine matrix elements stand for $^3P_0$, $^3P_1$, and $^1P_1$, respectively.

\begin{figure*}[h]
	\centering
		\includegraphics[width=14cm]{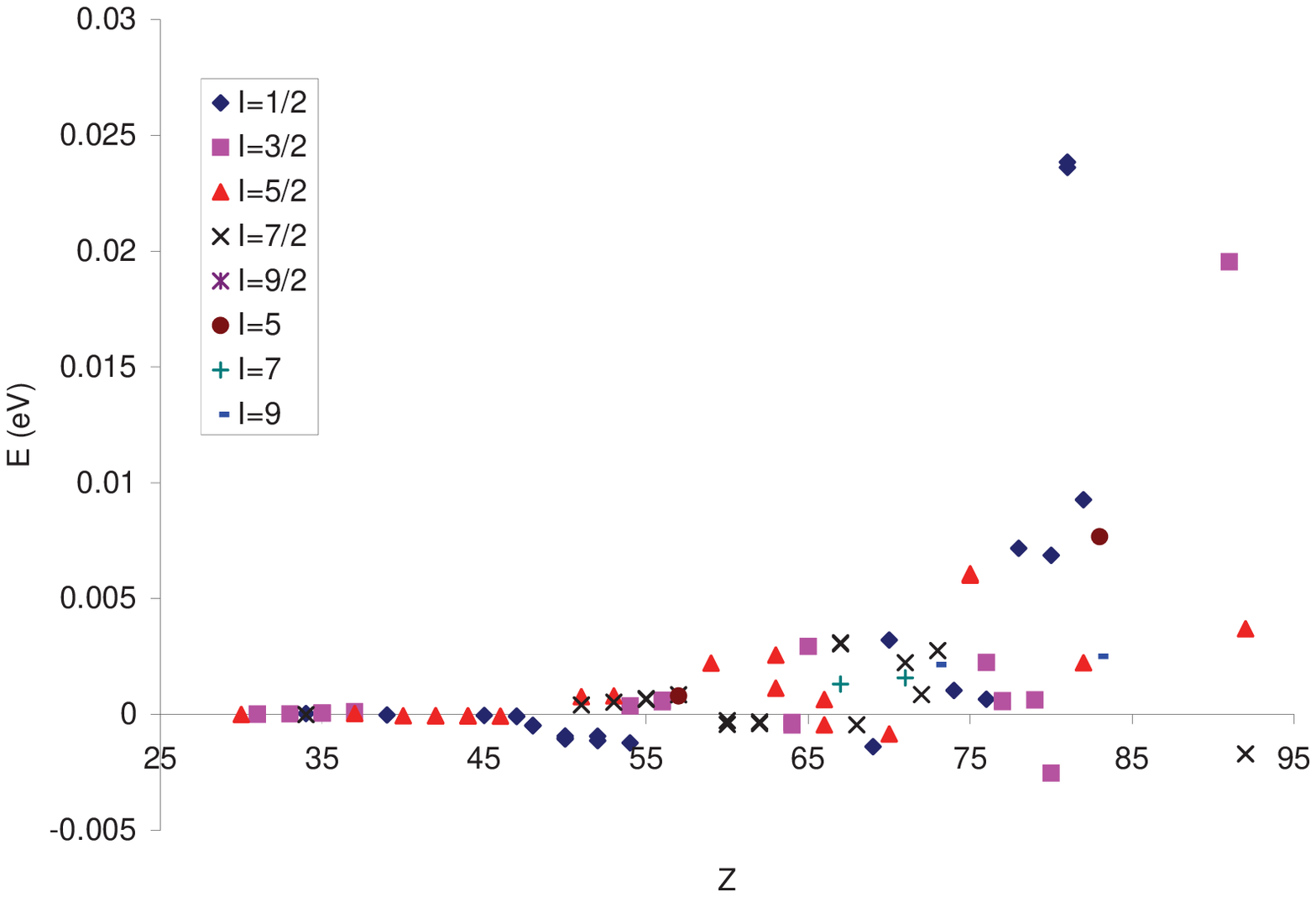}
	\caption{Influence of the hyperfine interaction on the $4s4p\ ^3P_1 \--\ ^3P_0$ energy separation, as a function of the nuclear spin $I$ and the atomic number $Z$. The quantity $E = \Delta E_{\textrm{hf}}-\Delta E_0$ is the contribution of the hyperfine interaction to the fine structure splitting $\Delta E_0$. The symbols represent values for the differente nuclear spins; some elements have several isotopes with identical spins but different $\mu_I$ values.}
	\label{fig2}
\end{figure*}

\begin{figure*}[h]
	\centering
		\includegraphics[width=14cm]{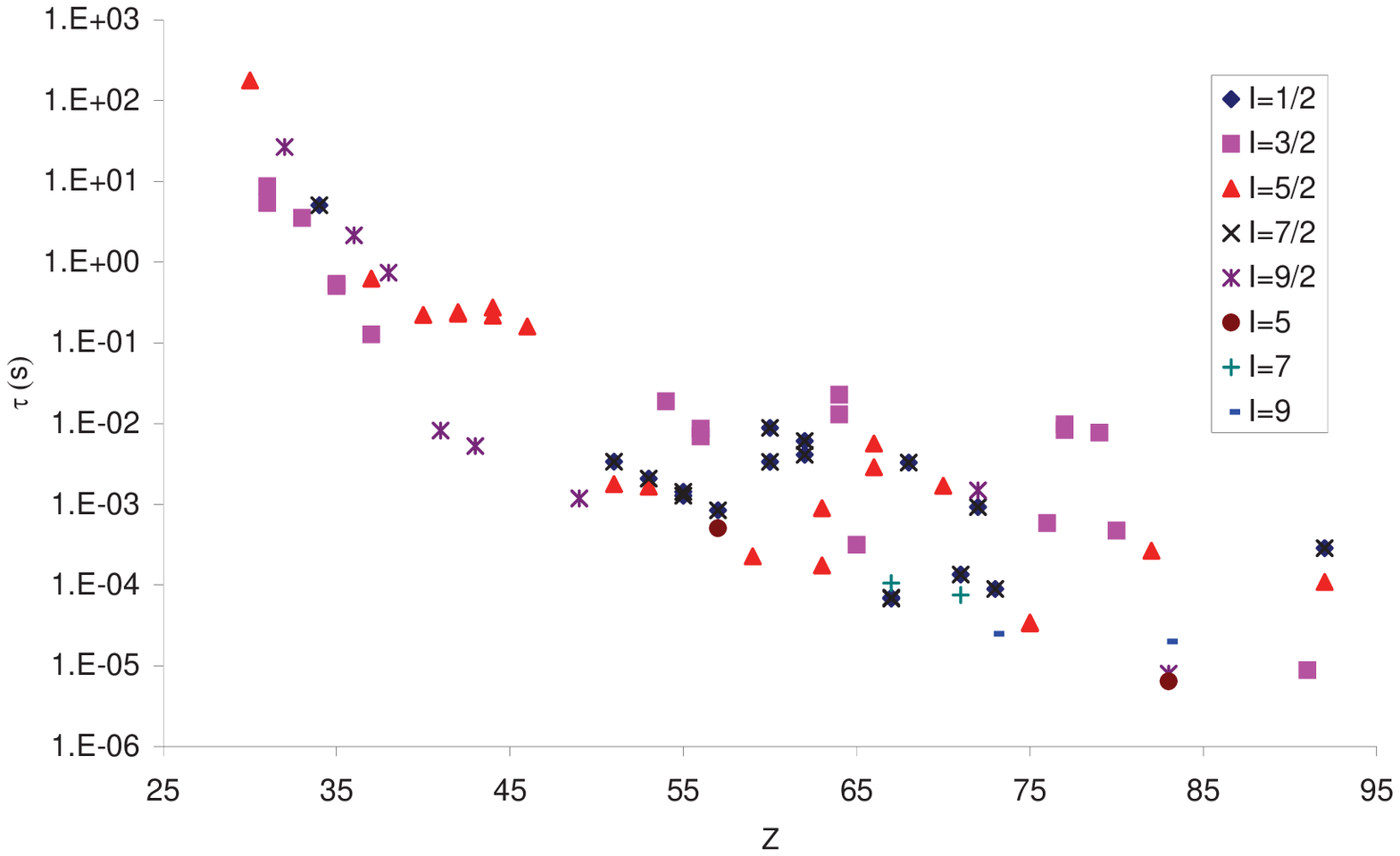}
	\caption{Influence of the hyperfine interaction on the lifetime of the $4s4p\ ^3P_0$ level, as a function of the nuclear spin $I$ and the atomic number $Z$. The unperturbed lifetime is, to a very good approximation, infinite.}
	\label{fig3}
\end{figure*}

In Table~\ref{lifetimes} are presented, for all possible
values of $I$, $Z$, and $A$, the
unperturbed separation energies $\Delta E_0=E_{^3P_1}-E_{^3P_0}$, the hyperfine-affected separation energy $\Delta E_{\textrm{hf}}$, 
the $4s4p\ ^3P_1$ and $4s4p\ ^1P_1$ levels lifetime values, $\tau_1$ and $\tau_2$, respectively, which are not affected, within
the precision shown, by the hyperfine interaction, and finally the perturbed lifetime $\tau_0$ of the $4s4p\ ^3P_0$ level. As we pointed out before, the unperturbed value of $\tau_0$ is assumed to be infinite.

In Fig.~\ref{fig2} is plotted the difference $E=\Delta E_{\textrm{hf}}-\Delta E_0$ as a function of $Z$ for the
different possible values of nuclear spin.  The influence of the hyperfine
interaction on this energy is shown to increase slowly with $Z$, the increase becoming more rapid for $Z\gtrapprox 60$.

In Fig.~\ref{fig3} is plotted the perturbed $4s4p\ ^3P_0$
level lifetime $\tau_0$, for the different nuclear spin values $I$, as a function of
$Z$.  One can easily conclude that the opening of a new channel for the decay
of the $4s4p\ ^3P_0$ level has a dramatic effect on its lifetime.

After submitting this paper, it was brought to our attention the work by Liu \textit{et al}~\cite{liu06},
which contains independent calculations of the hyperfine quenched lifetime of the $^3P_0$ level in several
Zn-like ions. Our results for this lifetime are three times higher than the values found by Liu \textit{et al} for
$Z=30$ and  1.5 higher at $Z=47$. To look for the origin of this discrepancy we used the
$^3P_1 \rightarrow ^1S_0$ and $^1P_1 \rightarrow ^1S_0$ transition energies and probabilities of Liu \textit{et al} 
to diagonalize the matrix in eq. \ref{eq:H} and obtained results very close to our own calculations. As we do not know 
the values of the hyperfine matrix elements calculated by Liu \textit{et al}, the reasons for this discrepancy remain unknown.

One of the most interesting practical implication of
these calculations comes from the relationship between the $4s4p\
^3P_0 - 4s4p\ ^3P_1$ levels energy separation and the $4s4p\
^3P_0$ level lifetime (Eq.~\ref{eq:H}).
As referred in ~\cite{jm2} this energy separation can be estimated from a measurement of the hyperfine-quenched $4s4p\ ^3P_0$ lifetime of
Zn-like ions with nuclear spin $I \neq 0$.  This method
has been demonstrated  for  heliumlike Ni$^{26+}$ \cite{dun1}, Ag$^{45+}$ \cite{rm1}, Gd$^{62+}$ \cite{indelicato92} and  Au$^{77+}$ \cite{toleikis04}.
In the Zn-like ions case, as in the corresponding Be-like and Mg-like ions, 
the situations is different, because, even for the highest $Z$ values, the lifetimes
involved are much longer than in heavy heliumlike ions.
However, measurements of Be-like hyperfine quenched transition rates have been performed from astrophysical sources~\cite{brage02} and the hyperfine quenching of the Zn-like $^{195}$Pt$^{48+}$ ion was observed in the TSR heavy-ion storage-ring~\cite{schippers05}. 

The continuous progress in storage rings, ion sources and ion traps leads us to believe that
lifetimes between 0.1 s and 10 $\mu$s
could be measured,  with some accuracy, by directly looking at the light
emitted by the ions as a function of time, after the trap has been loaded.
Very good vacuum inside the trap is needed, of course, if
long lifetimes are to be measured.
It remains to be demonstrated that this method is experimentally feasible.
Such experiments would be able to provide, for
different isotopes, the unperturbed energy
separation, because nuclear magnetic moments are well known.
This would be an interesting way to test our
relativistic calculations.


\begin{table*}[h]
\begin{center}
\caption{Hyperfine Matrix elements $W_{i,j}$ \ in eV. The indexes 0, 1, and 2 stand for $^3P_0$, $^3P_1$, and $^1P_1$ respectively. $I$ is the nuclear spin and $\mu$ the nuclear magnetic moment in nuclear magneton units.}
\label{matr_el}

\begin{tabular}{cccdddddd}\hline \hline
$I$	&	$Z$	&	$A$	&	\multicolumn{1}{c}{$W_{1,1}$}	&	\multicolumn{1}{c}{$W_{2,2}$}	&	\multicolumn{1}{c}{$W_{0,1}$}	&	\multicolumn{1}{c}{$W_{0,2}$}	&	\multicolumn{1}{c}{$W_{1,2}$}	&	\multicolumn{1}{c}{$\mu_I$}	\\
\hline
1/2	&	34	&	77	&	-3.312\times 10^{-05}	&	-3.245\times 10^{-06}	&	-2.382\times 10^{-05}	&	 1.708\times 10^{-05}	&	 3.268\times 10^{-05}	&	0.535042	\\
	  &	39	&	89	&	 2.697\times 10^{-05}	&	 4.139\times 10^{-07}	&	 1.670\times 10^{-05}	&	-1.055\times 10^{-05}	&	-2.199\times 10^{-05}	&	-0.137415	\\
	  &	45	&	103	&	 4.648\times 10^{-05}	&	-4.491\times 10^{-06}	&	 2.564\times 10^{-05}	&	-1.321\times 10^{-05}	&	-2.923\times 10^{-05}	&	-0.0884	\\
	  &	47	&	107	&	 7.878\times 10^{-05}	&	-1.013\times 10^{-05}	&	4.218\times 10^{-05}	&	-2.011 \times 10^{-05}	&	-4.507\times 10^{-05}	&	-0.11368	\\
	  &	47	&	109	&	9.057\times 10^{-05}	&	-1.164\times 10^{-05}	&	4.849\times 10^{-05}	&	-2.311\times 10^{-05}	&	-5.182\times 10^{-05}	&	-0.130691	\\
	  &	48	&	111	&	4.696\times 10^{-04}	&	-6.718\times 10^{-05}	&	2.480\times 10^{-04}	&	-1.136\times 10^{-04}	&	-2.560\times 10^{-04}	&	-0.594886	\\
  	&	48	&	113	&	4.912\times 10^{-04}	&	-7.027\times 10^{-05}	&	2.594\times 10^{-04}	&	-1.188\times 10^{-04}	&	-2.678\times 10^{-04}	&	-0.622301	\\
  	&	50	&	115	&	9.289\times 10^{-04}	&	-1.573\times 10^{-04}	&	4.788\times 10^{-04}	&	-2.018\times 10^{-04}	&	-4.591\times 10^{-04}	&	-0.91883	\\
  	&	50	&	117	&	1.012\times 10^{-03}	&	-1.714\times 10^{-04}	&	5.216\times 10^{-04}	&	-2.199\times 10^{-04}	&	-5.002\times 10^{-04}	&	-1.00104	\\
  	&	50	&	119	&	1.059\times 10^{-03}	&	-1.793\times 10^{-04}	&	5.457\times 10^{-04}	&	-2.301\times 10^{-04}	&	-5.233\times 10^{-04}	&	-1.04728	\\
  	&	52	&	123	&	9.385\times 10^{-04}	&	-1.801\times 10^{-04}	&	4.734\times 10^{-04}	&	-1.834\times 10^{-04}	&	-4.199\times 10^{-04}	&	-0.736948	\\
  	&	52	&	125	&	1.132\times 10^{-03}	&	-2.172\times 10^{-04}	&	5.708\times 10^{-04}	&	-2.212\times 10^{-04}	&	-5.063\times 10^{-04}	&	-0.888505	\\
  	&	54	&	129	&	1.232\times 10^{-03}	&	-2.602\times 10^{-04}	&	6.102\times 10^{-04}	&	-2.170\times 10^{-04}	&	-4.984\times 10^{-04}	&	-0.777976	\\
  	&	69	&	169	&	1.391\times 10^{-03}	&	-3.983\times 10^{-04}	&	6.364\times 10^{-04}	&	-1.217\times 10^{-04}	&	-2.710\times 10^{-04}	&	-0.231	\\
  	&	70	&	171	&	-3.210\times 10^{-03}	&	9.272\times 10^{-04}	&	-1.463\times 10^{-03}	&	2.694\times 10^{-04}	&	5.974\times 10^{-04}	&	0.49367	\\
  	&	74	&	183	&	-1.035\times 10^{-03}	&	3.081\times 10^{-04}	&	-4.659\times 10^{-04}	&	7.391\times 10^{-05}	&	1.610\times 10^{-04}	&	0.117785	\\
  	&	76	&	187	&	-6.581\times 10^{-04}	&	1.982\times 10^{-04}	&	-2.942\times 10^{-04}	&	4.344\times 10^{-05}	&	9.378\times 10^{-05}	&	0.064652	\\
  	&	78	&	195	&	-7.176\times 10^{-03}	&	2.186\times 10^{-03}	&	-3.189\times 10^{-03}	&	4.387\times 10^{-04}	&	9.383\times 10^{-04}	&	0.60952	\\
	  &	80	&	199	&	-6.867\times 10^{-03}	&	2.112\times 10^{-03}	&	-3.030\times 10^{-03}	&	3.892\times 10^{-04}	&	8.250\times 10^{-04}	&	0.505885	\\
	  &	81	&	203	&	-2.364\times 10^{-02}	&	7.305\times 10^{-03}	&	-1.040\times 10^{-02}	&	1.290\times 10^{-03}	&	2.724\times 10^{-03}	&	1.622258	\\
  	&	81	&	205	&	-2.387\times 10^{-02}	&	7.377\times 10^{-03}	&	-1.050\times 10^{-02}	&	1.303\times 10^{-03}	&	2.751\times 10^{-03}	&	1.638215	\\
  	&	82	&	207	&	-9.267\times 10^{-03}	&	2.876\times 10^{-03}	&	-4.060\times 10^{-03}	&	4.873\times 10^{-04}	&	1.024\times 10^{-03}	&	0.592583	\\
3/2	&	31	&	69	&	-1.363\times 10^{-05}	&	-1.123\times 10^{-06}	&	-2.652\times 10^{-05}	&	2.024\times 10^{-05}	&	1.535\times 10^{-05}	&	2.01659	\\
	&	31	&	71	&	-1.732\times 10^{-05}	&	-1.427\times 10^{-06}	&	-3.370\times 10^{-05}	&	2.572\times 10^{-05}	&	1.950\times 10^{-05}	&	2.56227	\\
	&	33	&	75	&	-2.179\times 10^{-05}	&	-2.286\times 10^{-06}	&	-3.670\times 10^{-05}	&	2.687\times 10^{-05}	&	2.235\times 10^{-05}	&	1.43948	\\
	&	35	&	79	&	-5.715\times 10^{-05}	&	-4.904\times 10^{-06}	&	-8.849\times 10^{-05}	&	6.206\times 10^{-05}	&	5.431\times 10^{-05}	&	2.1064	\\
	&	35	&	81	&	-6.161\times 10^{-05}	&	-5.286\times 10^{-06}	&	-9.538\times 10^{-05}	&	6.690\times 10^{-05}	&	5.855\times 10^{-05}	&	2.270562	\\
	&	37	&	87	&	-1.198\times 10^{-04}	&	-6.370\times 10^{-06}	&	-1.745\times 10^{-04}	&	1.165\times 10^{-04}	&	1.056\times 10^{-04}	&	2.75131	\\
	&	54	&	131	&	-3.649\times 10^{-04}	&	7.708\times 10^{-05}	&	-4.042\times 10^{-04}	&	1.438\times 10^{-04}	&	1.477\times 10^{-04}	&	0.6915	\\
	&	56	&	135	&	-5.429\times 10^{-04}	&	1.235\times 10^{-04}	&	-5.920\times 10^{-04}	&	1.932\times 10^{-04}	&	1.987\times 10^{-04}	&	0.837953	\\
	&	56	&	137	&	-6.073\times 10^{-04}	&	1.382\times 10^{-04}	&	-6.622\times 10^{-04}	&	2.161\times 10^{-04}	&	2.222\times 10^{-04}	&	0.937365	\\
	&	64	&	155	&	3.449\times 10^{-04}	&	-9.328\times 10^{-05}	&	3.595\times 10^{-04}	&	-8.381\times 10^{-05}	&	-8.504\times 10^{-05}	&	-0.2572	\\
	&	64	&	157	&	4.556\times 10^{-04}	&	-1.232\times 10^{-04}	&	4.750\times 10^{-04}	&	-1.107\times 10^{-04}	&	-1.123\times 10^{-04}	&	-0.3398	\\
	&	65	&	159	&	-2.936\times 10^{-03}	&	8.052\times 10^{-04}	&	-3.049\times 10^{-03}	&	6.825\times 10^{-04}	&	6.901\times 10^{-04}	&	2.014	\\
	&	76	&	189	&	-2.239\times 10^{-03}	&	6.745\times 10^{-04}	&	-2.238\times 10^{-03}	&	3.305\times 10^{-04}	&	3.191\times 10^{-04}	&	0.659933	\\
	&	77	&	191	&	-5.500\times 10^{-04}	&	1.666\times 10^{-04}	&	-5.481\times 10^{-04}	&	7.811\times 10^{-05}	&	7.506\times 10^{-05}	&	0.1507	\\
	&	77	&	193	&	-5.974\times 10^{-04}	&	1.810\times 10^{-04}	&	-5.954\times 10^{-04}	&	8.484\times 10^{-05}	&	8.154\times 10^{-05}	&	0.1637	\\
	&	79	&	197	&	-6.243\times 10^{-04}	&	1.911\times 10^{-04}	&	-6.182\times 10^{-04}	&	8.215\times 10^{-05}	&	7.824\times 10^{-05}	&	0.148158	\\
	&	80	&	201	&	2.535\times 10^{-03}	&	-7.798\times 10^{-04}	&	2.501\times 10^{-03}	&	-3.212\times 10^{-04}	&	-3.045\times 10^{-04}	&	-0.560226	\\
	&	91	&	231	&	-1.960\times 10^{-02}	&	6.265\times 10^{-03}	&	-1.845\times 10^{-02}	&	1.656\times 10^{-03}	&	1.506\times 10^{-03}	&	2.01	\\
5/2	&	30	&	67	&	-1.975\times 10^{-06}	&	-3.436\times 10^{-08}	&	-6.734\times 10^{-06}	&	5.115\times 10^{-06}	&	2.303\times 10^{-06}	&	0.875479	\\
	&	37	&	85	&	-3.537\times 10^{-05}	&	-1.880\times 10^{-06}	&	-7.865\times 10^{-05}	&	5.252\times 10^{-05}	&	3.118\times 10^{-05}	&	1.353352	\\
	&	40	&	91	&	6.158\times 10^{-05}	&	-2.554\times 10^{-07}	&	1.273\times 10^{-04}	&	-7.804\times 10^{-05}	&	-4.821\times 10^{-05}	&	-1.30362	\\
	&	42	&	95	&	6.081\times 10^{-05}	&	-2.597\times 10^{-06}	&	1.207\times 10^{-04}	&	-6.931\times 10^{-05}	&	-4.375\times 10^{-05}	&	-0.9142	\\
	&	42	&	97	&	6.209\times 10^{-05}	&	-2.652\times 10^{-06}	&	1.233\times 10^{-04}	&	-7.078\times 10^{-05}	&	-4.467\times 10^{-05}	&	-0.9335	\\
	&	44	&	99	&	5.826\times 10^{-05}	&	-4.623\times 10^{-06}	&	1.116\times 10^{-04}	&	-5.968\times 10^{-05}	&	-3.835\times 10^{-05}	&	-0.6413	\\
	&	44	&	101	&	6.530\times 10^{-05}	&	-5.182\times 10^{-06}	&	1.251\times 10^{-04}	&	-6.689\times 10^{-05}	&	-4.298\times 10^{-05}	&	-0.7188	\\
	&	46	&	105	&	7.773\times 10^{-05}	&	-8.788\times 10^{-06}	&	1.442\times 10^{-04}	&	-7.151\times 10^{-05}	&	-4.664\times 10^{-05}	&	-0.642	\\
	&	51	&	121	&	-7.648\times 10^{-04}	&	1.385\times 10^{-04}	&	-1.332\times 10^{-03}	&	5.382\times 10^{-04}	&	3.597\times 10^{-04}	&	3.3634	\\
	&	53	&	127	&	-8.004\times 10^{-04}	&	1.617\times 10^{-04}	&	-1.366\times 10^{-03}	&	5.072\times 10^{-04}	&	3.406\times 10^{-04}	&	2.813273	\\
	&	59	&	141	&	-2.218\times 10^{-03}	&	5.482\times 10^{-04}	&	-3.622\times 10^{-03}	&	1.040\times 10^{-03}	&	6.985\times 10^{-04}	&	4.2754	\\
	&	63	&	151	&	-2.567\times 10^{-03}	&	6.842\times 10^{-04}	&	-4.106\times 10^{-03}	&	9.970\times 10^{-04}	&	6.642\times 10^{-04}	&	3.4717	\\
	&	63	&	153	&	-1.133\times 10^{-03}	&	3.020\times 10^{-04}	&	-1.812\times 10^{-03}	&	4.401\times 10^{-04}	&	2.932\times 10^{-04}	&	1.5324	\\
	&	66	&	161	&	4.556\times 10^{-04}	&	-1.264\times 10^{-04}	&	7.196\times 10^{-04}	&	-1.548\times 10^{-04}	&	-1.021\times 10^{-04}	&	-0.4803	\\
	&	66	&	163	&	-6.384\times 10^{-04}	&	1.771\times 10^{-04}	&	-1.008\times 10^{-03}	&	2.169\times 10^{-04}	&	1.431\times 10^{-04}	&	0.673	\\
	&	70	&	173	&	8.428\times 10^{-04}	&	-2.434\times 10^{-04}	&	1.312\times 10^{-03}	&	-2.416\times 10^{-04}	&	-1.568\times 10^{-04}	&	-0.648	\\
	&	75	&	185	&	-6.034\times 10^{-03}	&	1.807\times 10^{-03}	&	-9.246\times 10^{-03}	&	1.415\times 10^{-03}	&	8.981\times 10^{-04}	&	3.1871	\\
	&	75	&	187	&	-6.096\times 10^{-03}	&	1.826\times 10^{-03}	&	-9.341\times 10^{-03}	&	1.429\times 10^{-03}	&	9.073\times 10^{-04}	&	3.2197	\\
	&	82	&	205	&	-2.226\times 10^{-03}	&	6.908\times 10^{-04}	&	-3.331\times 10^{-03}	&	3.998\times 10^{-04}	&	2.460\times 10^{-04}	&	0.7117	\\
	&	92	&	233	&	1.702\times 10^{-03}	&	-5.451\times 10^{-04}	&	3.263\times 10^{-03}	&	-2.839\times 10^{-04}	&	-1.256\times 10^{-04}	&	0.59	\\
	\hline
\end{tabular}
\end{center}
\end{table*}
\begin{table*}[h]
\begin{center}
\captcont{\emph{Continued}}
\begin{tabular}{cccdddddd}\hline \hline
$I$	&	$Z$	&	$A$	&	\multicolumn{1}{c}{$W_{1,1}$}	&	\multicolumn{1}{c}{$W_{2,2}$}	&	\multicolumn{1}{c}{$W_{0,1}$}	&	\multicolumn{1}{c}{$W_{0,2}$}	&	\multicolumn{1}{c}{$W_{1,2}$}	&	\multicolumn{1}{c}{$\mu_I$}	\\
\hline
7/2	&	34	&	79	&	9.003\times 10^{-06}	&	8.821\times 10^{-07}	&	2.967\times 10^{-05}	&	-2.128\times 10^{-05}	&	-8.883\times 10^{-06}	&	-1.018	\\
	&	51	&	123	&	-4.141\times 10^{-04}	&	7.500\times 10^{-05}	&	-9.673\times 10^{-04}	&	3.910\times 10^{-04}	&	1.948\times 10^{-04}	&	2.5498	\\
	&	53	&	129	&	-5.326\times 10^{-04}	&	1.076\times 10^{-04}	&	-1.220\times 10^{-03}	&	4.528\times 10^{-04}	&	2.267\times 10^{-04}	&	2.621	\\
	&	55	&	133	&	-6.480\times 10^{-04}	&	1.424\times 10^{-04}	&	-1.459\times 10^{-03}	&	4.970\times 10^{-04}	&	2.494\times 10^{-04}	&	2.582025	\\
	&	55	&	135	&	-6.857\times 10^{-04}	&	1.507\times 10^{-04}	&	-1.544\times 10^{-03}	&	5.260\times 10^{-04}	&	2.639\times 10^{-04}	&	2.7324	\\
	&	57	&	139	&	-8.527\times 10^{-04}	&	2.001\times 10^{-04}	&	-1.892\times 10^{-03}	&	5.915\times 10^{-04}	&	2.968\times 10^{-04}	&	2.783046	\\
	&	60	&	143	&	4.324\times 10^{-04}	&	-1.093\times 10^{-04}	&	9.424\times 10^{-04}	&	-2.593\times 10^{-04}	&	-1.296\times 10^{-04}	&	-1.065	\\
	&	60	&	145	&	2.664\times 10^{-04}	&	-6.731\times 10^{-05}	&	5.805\times 10^{-04}	&	-1.597\times 10^{-04}	&	-7.984\times 10^{-05}	&	-0.656	\\
	&	62	&	147	&	3.935\times 10^{-04}	&	-1.032\times 10^{-04}	&	8.483\times 10^{-04}	&	-2.147\times 10^{-04}	&	-1.069\times 10^{-04}	&	-0.812	\\
	&	62	&	149	&	3.235\times 10^{-04}	&	-8.486\times 10^{-05}	&	6.975\times 10^{-04}	&	-1.765\times 10^{-04}	&	-8.790\times 10^{-05}	&	-0.6677	\\
	&	67	&	163	&	-3.107\times 10^{-03}	&	8.721\times 10^{-04}	&	-6.561\times 10^{-03}	&	1.356\times 10^{-03}	&	6.644\times 10^{-04}	&	4.23	\\
	&	67	&	165	&	-3.063\times 10^{-03}	&	8.598\times 10^{-04}	&	-6.467\times 10^{-03}	&	1.337\times 10^{-03}	&	6.550\times 10^{-04}	&	4.17	\\
	&	68	&	167	&	4.484\times 10^{-04}	&	-1.271\times 10^{-04}	&	9.431\times 10^{-04}	&	-1.875\times 10^{-04}	&	-9.150\times 10^{-05}	&	-0.56385	\\
	&	71	&	175	&	-2.238\times 10^{-03}	&	6.514\times 10^{-04}	&	-4.657\times 10^{-03}	&	8.257\times 10^{-04}	&	3.979\times 10^{-04}	&	2.2323	\\
	&	72	&	177	&	-8.583\times 10^{-04}	&	2.519\times 10^{-04}	&	-1.781\times 10^{-03}	&	3.041\times 10^{-04}	&	1.459\times 10^{-04}	&	0.7935	\\
	&	73	&	181	&	-2.765\times 10^{-03}	&	8.176\times 10^{-04}	&	-5.722\times 10^{-03}	&	9.414\times 10^{-04}	&	4.495\times 10^{-04}	&	2.3705	\\
	&	92	&	235	&	-3.699\times 10^{-03}	&	1.185\times 10^{-03}	&	-5.287\times 10^{-03}	&	4.599\times 10^{-04}	&	2.731\times 10^{-04}	&	-0.38	\\
9/2	&	32	&	73	&	3.089\times 10^{-06}	&	3.176\times 10^{-07}	&	1.419\times 10^{-05}	&	-1.060\times 10^{-05}	&	-3.306\times 10^{-06}	&	-0.879468	\\
	&	36	&	83	&	1.124\times 10^{-05}	&	7.920\times 10^{-07}	&	4.326\times 10^{-05}	&	-2.963\times 10^{-05}	&	-1.029\times 10^{-05}	&	-0.970669	\\
	&	38	&	87	&	1.959\times 10^{-05}	&	6.773\times 10^{-07}	&	7.138\times 10^{-05}	&	-4.640\times 10^{-05}	&	-1.662\times 10^{-05}	&	-1.093603	\\
	&	41	&	93	&	-1.930\times 10^{-04}	&	4.552\times 10^{-06}	&	-6.573\times 10^{-04}	&	3.901\times 10^{-04}	&	1.449\times 10^{-04}	&	6.1705	\\
	&	43	&	99	&	-2.464\times 10^{-04}	&	1.512\times 10^{-05}	&	-8.076\times 10^{-04}	&	4.478\times 10^{-04}	&	1.696\times 10^{-04}	&	5.6847	\\
	&	49	&	113	&	-5.500\times 10^{-04}	&	8.617\times 10^{-05}	&	-1.648\times 10^{-03}	&	7.241\times 10^{-04}	&	2.856\times 10^{-04}	&	5.5289	\\
	&	49	&	115	&	-5.512\times 10^{-04}	&	8.636\times 10^{-05}	&	-1.651\times 10^{-03}	&	7.257\times 10^{-04}	&	2.862\times 10^{-04}	&	5.5408	\\
	&	72	&	179	&	5.392\times 10^{-04}	&	-1.582\times 10^{-04}	&	1.403\times 10^{-03}	&	-2.395\times 10^{-04}	&	-9.166\times 10^{-05}	&	-0.6409	\\
	&	83	&	209	&	-7.663\times 10^{-03}	&	2.388\times 10^{-03}	&	-1.922\times 10^{-02}	&	2.230\times 10^{-03}	&	8.126\times 10^{-04}	&	4.1103	\\
5	&	57	&	138	&	-7.965\times 10^{-04}	&	1.869\times 10^{-04}	&	-2.439\times 10^{-03}	&	7.625\times 10^{-04}	&	2.772\times 10^{-04}	&	3.713646	\\
	&	83	&	208	&	-7.773\times 10^{-03}	&	2.422\times 10^{-03}	&	-2.146\times 10^{-02}	&	2.491\times 10^{-03}	&	8.243\times 10^{-04}	&	4.633	\\
7	&	67	&	166	&	-1.322\times 10^{-03}	&	3.711\times 10^{-04}	&	-5.264\times 10^{-03}	&	1.088\times 10^{-03}	&	2.827\times 10^{-04}	&	3.6	\\
	&	71	&	176	&	-1.588\times 10^{-03}	&	4.624\times 10^{-04}	&	-6.233\times 10^{-03}	&	1.105\times 10^{-03}	&	2.824\times 10^{-04}	&	3.169	\\
9	&	73	&	180	&	-2.189\times 10^{-03}	&	6.472\times 10^{-04}	&	-1.083\times 10^{-02}	&	1.781\times 10^{-03}	&	3.558\times 10^{-04}	&	4.825	\\
	&	83	&	210	&	-2.545\times 10^{-03}	&	7.929\times 10^{-04}	&	-1.217\times 10^{-02}	&	1.412\times 10^{-03}	&	2.699\times 10^{-04}	&	2.73	\\
\hline
\end{tabular}
\end{center}
\end{table*}


\begin{table*}[h]
\begin{center}
\caption{Influence of the hyperfine interaction on the $4s4p\ ^3P_1 -4 s4p\ ^3P_0$ energy separation and on the lifetime of the $^3P_0$ level, as a function of the nuclear spin $I$ and the atomic number $Z$. $\Delta E_0$ is the unperturbed energy separation (in eV), and $\Delta E_{hf}$ is the perturbed energy (in eV) when the hyperfine interaction is taken into account (the 5 digits do not necessarily represent the accuracy of the calculation - they are intended to show the effect at low $Z$).
 $\tau_0$, $\tau_1$ and $\tau_2$ represent the perturbed lifetimes (in $s$) of $4s4p\ ^3P_0$, $^3P_1$ and $^1P_1$ levels respectively.}
\label{lifetimes}
\begin{tabular}{cccddddd}\hline \hline
I	&	Z	&	A	&	\multicolumn{1}{c}{$\Delta E_0$}	&	\multicolumn{1}{c}{$\Delta E_{hf}$}	&	\multicolumn{1}{c}{$\tau_1$}	&	\multicolumn{1}{c}{$\tau_2$}	&	\multicolumn{1}{c}{$\tau_0$}\\
\hline
1/2	&	34	&	77	&	0.20461	&	0.20458	&	2.139\times 10^{-07}	&	1.340\times 10^{-10}	&	7.847\times 10^{+00}	\\
	&	39	&	89	&	0.61806	&	0.61809	&	1.393\times 10^{-08}	&	5.025\times 10^{-11}	&	1.317\times 10^{+01}	\\
	&	45	&	103	&	1.41748	&	1.41753	&	1.976\times 10^{-09}	&	2.375\times 10^{-11}	&	5.111\times 10^{+00}	\\
	&	47	&	107	&	1.74706	&	1.74714	&	1.221\times 10^{-09}	&	1.923\times 10^{-11}	&	1.847\times 10^{+00}	\\
	&	47	&	109	&	1.74706	&	1.74715	&	1.221\times 10^{-09}	&	1.923\times 10^{-11}	&	1.397\times 10^{+00}	\\
	&	48	&	111	&	1.92132	&	1.92179	&	9.804\times 10^{-10}	&	1.739\times 10^{-11}	&	5.286\times 10^{-02}	\\
	&	48	&	113	&	1.92132	&	1.92181	&	9.804\times 10^{-10}	&	1.739\times 10^{-11}	&	4.831\times 10^{-02}	\\
	&	50	&	115	&	2.28581	&	2.28674	&	6.623\times 10^{-10}	&	1.433\times 10^{-11}	&	1.391\times 10^{-02}	\\
	&	50	&	117	&	2.28581	&	2.28682	&	6.623\times 10^{-10}	&	1.433\times 10^{-11}	&	1.172\times 10^{-02}	\\
	&	50	&	119	&	2.28581	&	2.28687	&	6.623\times 10^{-10}	&	1.433\times 10^{-11}	&	1.071\times 10^{-02}	\\
	&	52	&	123	&	2.66778	&	2.66872	&	4.673\times 10^{-10}	&	1.185\times 10^{-11}	&	1.397\times 10^{-02}	\\
	&	52	&	125	&	2.66778	&	2.66891	&	4.673\times 10^{-10}	&	1.185\times 10^{-11}	&	9.609\times 10^{-03}	\\
	&	54	&	129	&	3.06295	&	3.06418	&	3.425\times 10^{-10}	&	9.804\times 10^{-12}	&	8.274\times 10^{-03}	\\
	&	69	&	169	&	6.17717	&	6.17856	&	7.752\times 10^{-11}	&	2.398\times 10^{-12}	&	7.234\times 10^{-03}	\\
	&	70	&	171	&	6.38647	&	6.38326	&	7.246\times 10^{-11}	&	2.174\times 10^{-12}	&	1.367\times 10^{-03}	\\
	&	74	&	183	&	7.22323	&	7.22219	&	5.650\times 10^{-11}	&	1.464\times 10^{-12}	&	1.353\times 10^{-02}	\\
	&	76	&	187	&	7.64172	&	7.64106	&	5.051\times 10^{-11}	&	1.196\times 10^{-12}	&	3.400\times 10^{-02}	\\
	&	78	&	195	&	8.06074	&	8.05357	&	4.545\times 10^{-11}	&	9.804\times 10^{-13}	&	2.895\times 10^{-04}	\\
	&	80	&	199	&	8.48051	&	8.47364	&	4.115\times 10^{-11}	&	7.937\times 10^{-13}	&	3.215\times 10^{-04}	\\
	&	81	&	203	&	8.69084	&	8.66722	&	3.922\times 10^{-11}	&	7.194\times 10^{-13}	&	2.725\times 10^{-05}	\\
	&	81	&	205	&	8.69084	&	8.66699	&	3.922\times 10^{-11}	&	7.194\times 10^{-13}	&	2.672\times 10^{-05}	\\
	&	82	&	207	&	8.90152	&	8.89226	&	3.745\times 10^{-11}	&	6.494\times 10^{-13}	&	1.794\times 10^{-04}	\\
3/2	&	31	&	69	&	0.06865	&	0.06864	&	4.446\times 10^{-06}	&	4.528\times 10^{-10}	&	8.711\times 10^{+00}	\\
	&	31	&	71	&	0.06865	&	0.06863	&	4.446\times 10^{-06}	&	4.528\times 10^{-10}	&	5.395\times 10^{+00}	\\
	&	33	&	75	&	0.14992	&	0.14990	&	4.831\times 10^{-07}	&	1.808\times 10^{-10}	&	3.524\times 10^{+00}	\\
	&	35	&	79	&	0.26849	&	0.26843	&	1.071\times 10^{-07}	&	1.044\times 10^{-10}	&	5.357\times 10^{-01}	\\
	&	35	&	81	&	0.26849	&	0.26843	&	1.071\times 10^{-07}	&	1.044\times 10^{-10}	&	4.610\times 10^{-01}	\\
	&	37	&	87	&	0.42429	&	0.42417	&	3.448\times 10^{-08}	&	6.944\times 10^{-11}	&	1.270\times 10^{-01}	\\
	&	54	&	131	&	3.06295	&	3.06259	&	3.425\times 10^{-10}	&	9.804\times 10^{-12}	&	1.883\times 10^{-02}	\\
	&	56	&	135	&	3.46760	&	3.46706	&	2.611\times 10^{-10}	&	8.197\times 10^{-12}	&	8.660\times 10^{-03}	\\
	&	56	&	137	&	3.46760	&	3.46699	&	2.611\times 10^{-10}	&	8.197\times 10^{-12}	&	6.920\times 10^{-03}	\\
	&	64	&	155	&	5.12988	&	5.13022	&	1.130\times 10^{-10}	&	3.876\times 10^{-12}	&	2.277\times 10^{-02}	\\
	&	64	&	157	&	5.12988	&	5.13034	&	1.130\times 10^{-10}	&	3.876\times 10^{-12}	&	1.304\times 10^{-02}	\\
	&	65	&	159	&	5.33938	&	5.33645	&	1.041\times 10^{-10}	&	3.534\times 10^{-12}	&	3.155\times 10^{-04}	\\
	&	76	&	189	&	7.64172	&	7.63948	&	5.051\times 10^{-11}	&	1.196\times 10^{-12}	&	5.872\times 10^{-04}	\\
	&	77	&	191	&	7.85115	&	7.85060	&	4.785\times 10^{-11}	&	1.081\times 10^{-12}	&	9.806\times 10^{-03}	\\
	&	77	&	193	&	7.85115	&	7.85055	&	4.785\times 10^{-11}	&	1.081\times 10^{-12}	&	8.310\times 10^{-03}	\\
	&	79	&	197	&	8.27049	&	8.26987	&	4.329\times 10^{-11}	&	8.850\times 10^{-13}	&	7.729\times 10^{-03}	\\
	&	80	&	201	&	8.48051	&	8.48305	&	4.115\times 10^{-11}	&	7.937\times 10^{-13}	&	4.730\times 10^{-04}	\\
	&	91	&	231	&	10.82070	&	10.80116	&	2.591\times 10^{-11}	&	2.538\times 10^{-13}	&	8.872\times 10^{-06}	\\
5/2	&	30	&	67	&	0.04546	&	0.04546	&	1.834\times 10^{-05}	&	1.371\times 10^{-09}	&	1.771\times 10^{+02}	\\
	&	37	&	85	&	0.42429	&	0.42425	&	3.448\times 10^{-08}	&	6.944\times 10^{-11}	&	6.251\times 10^{-01}	\\
	&	40	&	91	&	0.72914	&	0.72920	&	9.434\times 10^{-09}	&	4.348\times 10^{-11}	&	2.221\times 10^{-01}	\\
	&	42	&	95	&	0.97889	&	0.97895	&	4.695\times 10^{-09}	&	3.344\times 10^{-11}	&	2.395\times 10^{-01}	\\
	&	42	&	97	&	0.97889	&	0.97895	&	4.695\times 10^{-09}	&	3.344\times 10^{-11}	&	2.297\times 10^{-01}	\\
	&	44	&	99	&	1.26339	&	1.26345	&	2.584\times 10^{-09}	&	2.646\times 10^{-11}	&	2.732\times 10^{-01}	\\
	&	44	&	101	&	1.26339	&	1.26346	&	2.584\times 10^{-09}	&	2.646\times 10^{-11}	&	2.175\times 10^{-01}	\\
	&	46	&	105	&	1.57892	&	1.57900	&	1.541\times 10^{-09}	&	2.132\times 10^{-11}	&	1.597\times 10^{-01}	\\
	&	51	&	121	&	2.47489	&	2.47413	&	5.525\times 10^{-10}	&	1.302\times 10^{-11}	&	1.779\times 10^{-03}	\\
	&	53	&	127	&	2.86396	&	2.86316	&	3.984\times 10^{-10}	&	1.079\times 10^{-11}	&	1.661\times 10^{-03}	\\
	&	59	&	141	&	4.08585	&	4.08364	&	1.828\times 10^{-10}	&	6.211\times 10^{-12}	&	2.274\times 10^{-04}	\\
	&	63	&	151	&	4.92049	&	4.91793	&	1.233\times 10^{-10}	&	4.274\times 10^{-12}	&	1.747\times 10^{-04}	\\
	&	63	&	153	&	4.92049	&	4.91936	&	1.233\times 10^{-10}	&	4.274\times 10^{-12}	&	8.977\times 10^{-04}	\\
	&	66	&	161	&	5.54887	&	5.54933	&	9.615\times 10^{-11}	&	3.205\times 10^{-12}	&	5.667\times 10^{-03}	\\
	&	66	&	163	&	5.54887	&	5.54823	&	9.615\times 10^{-11}	&	3.205\times 10^{-12}	&	2.885\times 10^{-03}	\\
	&	70	&	173	&	6.38647	&	6.38731	&	7.246\times 10^{-11}	&	2.174\times 10^{-12}	&	1.702\times 10^{-03}	\\
	\hline
\end{tabular}
\end{center}
\end{table*}


\begin{table*}[h]
\begin{center}
\captcont{\emph{Continued}}
\begin{tabular}{cccddddd}\hline \hline
I	&	Z	&	A	&	\multicolumn{1}{c}{$\Delta E_0$}	&	\multicolumn{1}{c}{$\Delta E_{hf}$}	&	\multicolumn{1}{c}{$\tau_1$}	&	\multicolumn{1}{c}{$\tau_2$}	&	\multicolumn{1}{c}{$\tau_0$}\\
\hline
5/2	&	75	&	185	&	7.43246	&	7.42645	&	5.348\times 10^{-11}	&	1.325\times 10^{-12}	&	3.433\times 10^{-05}	\\
	&	75	&	187	&	7.43246	&	7.42639	&	5.348\times 10^{-11}	&	1.325\times 10^{-12}	&	3.364\times 10^{-05}	\\
	&	82	&	205	&	8.90152	&	8.89930	&	3.745\times 10^{-11}	&	6.494\times 10^{-13}	&	2.670\times 10^{-04}	\\
	&	92	&	233	&	11.03725	&	11.03356	&	2.506\times 10^{-11}	&	2.288\times 10^{-13}	&	1.090\times 10^{-04}	\\
7/2	&	34	&	79	&	0.20461	&	0.20462	&	2.139\times 10^{-07}	&	1.340\times 10^{-10}	&	5.060\times 10^{+00}	\\
	&	51	&	123	&	2.47489	&	2.47448	&	5.525\times 10^{-10}	&	1.302\times 10^{-11}	&	3.371\times 10^{-03}	\\
	&	53	&	129	&	2.86396	&	2.86343	&	3.984\times 10^{-10}	&	1.079\times 10^{-11}	&	2.084\times 10^{-03}	\\
	&	55	&	133	&	3.26430	&	3.26365	&	2.976\times 10^{-10}	&	8.929\times 10^{-12}	&	1.435\times 10^{-03}	\\
	&	55	&	135	&	3.26430	&	3.26362	&	2.976\times 10^{-10}	&	8.929\times 10^{-12}	&	1.281\times 10^{-03}	\\
	&	57	&	139	&	3.67249	&	3.67164	&	2.304\times 10^{-10}	&	7.463\times 10^{-12}	&	8.424\times 10^{-04}	\\
	&	60	&	143	&	4.29381	&	4.29424	&	1.645\times 10^{-10}	&	5.650\times 10^{-12}	&	3.350\times 10^{-03}	\\
	&	60	&	145	&	4.29381	&	4.29408	&	1.645\times 10^{-10}	&	5.650\times 10^{-12}	&	8.830\times 10^{-03}	\\
	&	62	&	147	&	4.71127	&	4.71166	&	1.351\times 10^{-10}	&	4.695\times 10^{-12}	&	4.108\times 10^{-03}	\\
	&	62	&	149	&	4.71127	&	4.71159	&	1.351\times 10^{-10}	&	4.695\times 10^{-12}	&	6.075\times 10^{-03}	\\
	&	67	&	163	&	5.75836	&	5.75527	&	8.929\times 10^{-11}	&	2.915\times 10^{-12}	&	6.802\times 10^{-05}	\\
	&	67	&	165	&	5.75836	&	5.75531	&	8.929\times 10^{-11}	&	2.915\times 10^{-12}	&	6.999\times 10^{-05}	\\
	&	68	&	167	&	5.96779	&	5.96824	&	8.264\times 10^{-11}	&	2.646\times 10^{-12}	&	3.294\times 10^{-03}	\\
	&	71	&	175	&	6.59569	&	6.59346	&	6.757\times 10^{-11}	&	1.972\times 10^{-12}	&	1.350\times 10^{-04}	\\
	&	72	&	177	&	6.80490	&	6.80404	&	6.369\times 10^{-11}	&	1.786\times 10^{-12}	&	9.241\times 10^{-04}	\\
	&	73	&	181	&	7.01409	&	7.01133	&	5.988\times 10^{-11}	&	1.618\times 10^{-12}	&	8.953\times 10^{-05}	\\
	&	92	&	235	&	11.03725	&	11.03895	&	2.506\times 10^{-11}	&	2.288\times 10^{-13}	&	2.864\times 10^{-04}	\\
9/2	&	32	&	73	&	0.10442	&	0.10442	&	1.298\times 10^{-06}	&	2.644\times 10^{-10}	&	2.655\times 10^{+01}	\\
	&	36	&	83	&	0.34167	&	0.34168	&	5.882\times 10^{-08}	&	8.403\times 10^{-11}	&	2.142\times 10^{+00}	\\
	&	38	&	87	&	0.51641	&	0.51643	&	2.146\times 10^{-08}	&	5.882\times 10^{-11}	&	7.376\times 10^{-01}	\\
	&	41	&	93	&	0.84950	&	0.84931	&	6.536\times 10^{-09}	&	3.802\times 10^{-11}	&	8.194\times 10^{-03}	\\
	&	43	&	99	&	1.11697	&	1.11672	&	3.448\times 10^{-09}	&	2.967\times 10^{-11}	&	5.277\times 10^{-03}	\\
	&	49	&	113	&	2.10110	&	2.10055	&	8.000\times 10^{-10}	&	1.577\times 10^{-11}	&	1.184\times 10^{-03}	\\
	&	49	&	115	&	2.10110	&	2.10055	&	8.000\times 10^{-10}	&	1.577\times 10^{-11}	&	1.179\times 10^{-03}	\\
	&	72	&	179	&	6.80490	&	6.80544	&	6.369\times 10^{-11}	&	1.786\times 10^{-12}	&	1.491\times 10^{-03}	\\
	&	83	&	209	&	9.11260	&	9.10502	&	3.584\times 10^{-11}	&	5.848\times 10^{-13}	&	8.025\times 10^{-06}	\\
5	&	57	&	138	&	3.67249	&	3.67170	&	2.304\times 10^{-10}	&	7.463\times 10^{-12}	&	5.069\times 10^{-04}	\\
	&	83	&	208	&	9.11260	&	9.10493	&	3.584\times 10^{-11}	&	5.848\times 10^{-13}	&	6.433\times 10^{-06}	\\
7	&	67	&	166	&	5.75836	&	5.75705	&	8.929\times 10^{-11}	&	2.915\times 10^{-12}	&	1.057\times 10^{-04}	\\
	&	71	&	176	&	6.59569	&	6.59411	&	6.757\times 10^{-11}	&	1.972\times 10^{-12}	&	7.540\times 10^{-05}	\\
9	&	73	&	180	&	7.01409	&	7.01193	&	5.988\times 10^{-11}	&	1.618\times 10^{-12}	&	2.501\times 10^{-05}	\\
	&	83	&	210	&	9.11260	&	9.11009	&	3.584\times 10^{-11}	&	5.848\times 10^{-13}	&	2.003\times 10^{-05}	\\
	\hline
\end{tabular}
\end{center}
\end{table*}

\section*{Acknowledgments}

This research was partially supported by the FCT projects
POCTI/FAT/50356/2002 and POCTI/0303/2003  (Portugal),
financed by the European Community Fund FEDER, and by the French-Portuguese collaboration (PESSOA Program, Contract n$^{\circ}$ 10721NF). Laboratoire Kastler Brossel is Unit{\'e} Mixte de Recherche du CNRS
n$^{\circ}$ C8552.


\end{document}